\documentclass[twocolumn,pre,showpacs]{revtex4}
\usepackage{graphicx}
\usepackage{amsmath}
\usepackage{amsfonts}
\usepackage{mathtools}
\usepackage{subfigure}

\begin{document}
\title{Dispersion for Two Classes of Random Variables: General Theory and\\ Application to Inference of an External Ligand Concentration by a Cell}

\author{Andre C. Barato and Udo Seifert}
\affiliation{ II. Institut f\"ur Theoretische Physik, Universit\"at Stuttgart, 70550 Stuttgart, Germany}

\parskip 1mm
\newcommand{\e}{{\rm e}}
\def\d{{\rm d}}
\def\Ps{{P_{\scriptscriptstyle \hspace{-0.3mm} s}}}
\def\MF{{\mbox{\tiny \rm \hspace{-0.3mm} MF}}}
\def\i{\text{\scriptsize $\cal{I}$}} 

\begin{abstract}
We derive expressions for the dispersion for two classes of random variables in Markov processes.
Random variables like current and activity pertain to the first class, which is composed by random variables
that change whenever a jump in the stochastic trajectory occurs. The second class corresponds to the time the trajectory 
spends in a state (or cluster of states). While the expression for the first class follows straightforwardly from known results in the literature,
we show that a similar formalism can be used to derive an expression for the second class. As an application,
we use this formalism to analyze a cellular two-component network estimating an external ligand concentration.    
The uncertainty related to this external concentration is calculated by monitoring different random variables related 
to an internal protein. We show that, {\sl inter alia}, monitoring the time spent in the 
phosphorylated state of the protein leads to a finite uncertainty only if there is dissipation, whereas 
the uncertainty obtained from the activity of the transitions of the internal protein can reach the Berg and Purcell limit 
even in equilibrium. 
\end{abstract}
% Explanation of PACS numbers:
% 87.10.Vg: Biological information 
% 05.70.Ln: Nonequilibrium and irreversible thermodynamics
% 02.50.Ey: Stochastic processes 

\maketitle
%==========================================================================
\section{Introduction}
%==========================================================================

Markov processes are used to model a wide variety of physical, chemical and biological phenomena  \cite{vankampen,gardiner,bial12}. 
While calculating the mean of a random variable is often relatively straightforward, obtaining its dispersion can be much harder. A prominent 
example is the expression for the dispersion of the number of steps of a particle hopping in a one-dimensional lattice \cite{derr83}.      

Fluctuating currents are  important random variables in stochastic thermodynamics \cite{seif12}. If their mean is nonzero in the 
steady state, the system is out equilibrium. A standard method to calculate the dispersion of this random variable is to obtain
the scaled cumulant generating function as the maximum eigenvalue of the so-called modified generator \cite{lebo99}. However,
determining this eigenvalue as a function of the transition rates is typically difficult, even for systems with a small number of states.  

The problem of calculating the dispersion of fluctuating currents can be overcome with an elegant method introduced by Koza 
\cite{koza99}. With Koza's method, instead of calculating the maximum eigenvalue, the dispersion is obtained from coefficients 
of the characteristic polynomial associated with the modified generator. For instance, this method is useful to prove 
a bound on the minimal number of states of a one-dimensional lattice in terms of this dispersion \cite{koza02} (see \cite{kolo07,moff14} for discussions on the importance of this bound).
Recently, we have used this method to derive a bound on the minimal energetic cost of small uncertainty in biomolecular processes \cite{bara15}. 
Other references that use this method are \cite{chem08,wach14,alta15,bara15a} (see also \cite{flin08,flin10,rond09} for related methods that use perturbation theory). 

As we point out in the present paper, Koza's method can be used to calculate the dispersion of any random variable that changes whenever 
a jump occurs during a stochastic trajectory. Besides fluctuating currents, another random variable pertaining to this class is the 
dynamical activity, for which the total number of transitions is counted. This activity appears in fluctuation-dissipation 
relations out of equilibrium  \cite{baie09,baie13} and is studied in so-called dynamical phase transitions  \cite{leco07a,garr07,spec11a}.   

We show that a similar method can be used to calculate the dispersion of a second class of random variables. This class corresponds 
to variables counting the fraction of time a stochastic trajectory spends in a cluster of states. Our calculations lead to
an expression for the dispersion of this variable in terms of the coefficients of the characteristic polynomial associated with a discrete-time 
modified generator. 
 
The time spent in a state is a key random variable in the problem of a cell estimating an external ligand 
concentration by monitoring the time a receptor is in the bound state, i.e., occupied by an external ligand. The uncertainty on the estimated concentration 
sets a fundamental limit on the cell's ability to sense the external environment, as established in the seminal work of Berg and Purcell \cite{berg77} (see also \cite{bial05,endr08,endr09,mora10,gove12,kaiz14}).    

We analyze a four-state model for a two-component network that estimates an external concentration \cite{meht12,bara13b,bara14a,gove14,gove14a}. The two-component network is composed by the receptor and
an internal protein that can be phosphorylated due to ATP consumption. This model provides an example for which we can 
calculate the dispersion for three different random variables: current and activity from the first class, and time spent in the phosphorylated state from the second.
Specifically, we compare the uncertainty of the estimated external concentration that is obtained from these three different random variables and discuss its relation
with energy dissipation.  

The paper is organized as follows. The expression for the dispersion of the first class of random variables is provided in Sec. \ref{sec2}. In Sec. 
\ref{sec3}, we introduce a method to calculate the dispersion for the second class. Sec. \ref{sec4} contains the application to the problem of 
sensing an external concentration. We conclude in Sec. \ref{sec5}. In Appendices \ref{app1} and  \ref{app2}, we perform detailed derivations for
the first and second classes of random variables, respectively. The calculations of the four-state Markov process considered in Sec. \ref{sec4} are
explained in Appendix \ref{app3}.

%\begin{figure}
%\includegraphics[width=72mm]{./fig1new.eps}
%\vspace{-2mm}
%\caption{(Color online) Network and transition rates of a simple model of an internal process $Y$ coupled to an external one $X$.}
%\label{fig1} 
%\end{figure}

%==========================================================================
\section{Dispersion for current and activity}
%==========================================================================
\label{sec2}

The transition rate from state $i$ to state $j$ of a generic continuous-time Markov process with a finite number of states $N$
is denoted by $w_{ij}$. The time evolution of the occupation probability of state $i$ at time $\tau$ is determined by the master equation 
\begin{equation}
\frac{d}{d\tau}P_i(\tau)= \sum_{j\neq i}\left[P_j(\tau)w_{ji}-P_i(\tau)w_{ij}\right].
\end{equation}
For the discussion below it is convenient to write this equation in the from 
\begin{equation}
\frac{d}{d\tau}\mathbf{P}(\tau)= \mathcal{L}\mathbf{P}(\tau),
\end{equation}
where $\mathbf{P}(\tau)$ is the probability vector with $N$ components and $\mathcal{L}$ is the stochastic matrix with 
elements 
\begin{equation}
[\mathcal{L}]_{ij}\equiv\left\{\begin{array}{ll} 
 w_{ji} & \quad \textrm{if } i\neq j\\
 -\sum_kw_{ik} & \quad \textrm{if } i=j
\end{array}\right.\,.
\label{generator}
\end{equation}
Furthermore, the stationary probability of state $i$ is written as $P_i$. Two key quantities in this paper are the (stationary) average probability current from state $i$ to $j$ 
\begin{equation}
J_{ij}\equiv P_iw_{ij}-P_jw_{ji}
\end{equation}
and the average activity
\begin{equation}
A_{ij}\equiv P_iw_{ij}+P_jw_{ji}.
\end{equation}
This activity is the average number of transitions per time between the pair of states $ij$.

Quite generally, we consider a random variable $X_1$ that is a functional of the stochastic trajectory. This trajectory is a sequence of jumps during a time interval $t$,
which is assumed to be long, i.e., the expressions below are valid in the formal limit $t\to\infty$. The random variable $X_1$ is then characterized by the increments
$\theta_{ij}$: it increases by $\theta_{ij}$ whenever a jump from $i$ to $j$ occurs. 
If we choose the increments $\theta_{ij}=-\theta_{ji}=1$  and zero for any other pair of states, we obtain $\langle X_1\rangle/t= J_{ij}$, where here and in the following 
the brackets denotes an average over stochastic trajectories in the stationary state. On the other hand, the choice $\theta_{ij}=\theta_{ji}=1$ results in 
$\langle X_1\rangle/t= A_{ij}$. The ``velocity'' associated with the generic variable $X_1$ is
\begin{equation}
v_1\equiv \frac{\langle X_1\rangle}{t}.
\end{equation}
Furthermore, the dispersion is defined as 
\begin{equation}
D_1\equiv \frac{\langle X_1^2\rangle-\langle X_1\rangle^2}{2t}.
\end{equation}

An expression for these quantities in terms of the transition rates is obtained in the following way.
The scaled cumulant generating function associated with  $X_1$ is given by the maximum eigenvalue of the modified generator \cite{lebo99,touc09} 
\begin{equation}
[\mathcal{L}(z)]_{ij}\equiv\left\{\begin{array}{ll} 
 w_{ji}\exp(z \theta_{ji}) & \quad \textrm{if } i\neq j\\
 -\sum_kw_{ik} & \quad \textrm{if } i=j
\end{array}\right.\,.
\label{modgenerator}
\end{equation}
The characteristic polynomial of this matrix reads
\begin{equation}
\det\left(yI-\mathcal{L}(z)\right)\equiv \sum_{n=0}^{N}C_n(z)y^n,
\label{eqdet}
\end{equation}
where $C_n(z)$ are coefficients of the polynomial and $I$ is the identity matrix. It turns out that the velocity and dispersion can be written in terms of these coefficients and its derivatives at $z=0$. 
This procedure is very convenient since we can calculate the dispersion without knowing the full scaled cumulant generating function, which is a root of this polynomial. 
The specific expressions are 
\begin{equation}
v_1= -\frac{C_0'}{C_1},
\label{eqJC}
\end{equation}
and
\begin{equation}
D_1= -\frac{C_0''+2C_1'v_1+2C_2 v_1^2}{2C_1},
\label{eqDC}
\end{equation}
where primes denote derivatives with respect to $z$ and the lack of explicit $z$ dependence in the coefficients  denotes evaluation at $z=0$. 
These expressions are derived in appendix \ref{app1}. They have been obtained by Koza \cite{koza99} for the random variable $X_1$ being the current. 
We are simply pointing out here that they are also valid for any variable $X_1$ with generic increments $\theta_{ij}$, i.e., in particular also for the activity. 

%==========================================================================
\section{Dispersion for the time spent in a state}
%==========================================================================
\label{sec3}

The second class of random variables $X_2$ corresponds to the time the stochastic trajectory spends in a state or in a cluster of states.
The first moment of this random variable is easy to determine. For example, the average time spent in state $i$ is the stationary probability $P_i$ multiplied by
the total time, i.e.,  $\langle X_2\rangle= P_it$. Generally, we denote a cluster of states by $\Gamma$, with $\delta_{i,\Gamma}=1$ if state $i\in \Gamma$ and
$\delta_{i,\Gamma}=0$ otherwise. The ``velocity'' associated with $X_2$ is defined as
\begin{equation}
v_2\equiv\frac{\langle X_2\rangle}{t}= \sum_{i}P_i\delta_{i,\Gamma}.
\end{equation}
Furthermore, the dispersion reads
\begin{equation}
D_2\equiv \frac{\langle X_2^2\rangle-\langle X_2\rangle^2}{2t}.
\end{equation}

We discretize time in order to calculate this dispersion. The discrete-time stochastic matrix is
\begin{equation}
\mathcal{K}\equiv I+\epsilon \mathcal{L},
\label{discsto}
\end{equation}
where $\epsilon$ is the size of the time-interval. Since we are interested in the continuous-time limit, at the end of the calculation
the limit $\epsilon\to 0$ is taken. We define the modified matrix  
\begin{equation}
[\mathcal{K}(z)]_{ij}\equiv\left\{\begin{array}{ll} 
 \epsilon w_{ji}\exp(z \delta_{\Gamma,i}) & \quad \textrm{if } i\neq j\\
 \left(1-\epsilon\sum_kw_{ik}\right)\exp(z \delta_{\Gamma,i}) & \quad \textrm{if } i=j
\end{array}\right.\,.
\label{modgeneratordisc}
\end{equation}
The maximum eigenvalue of this modified matrix generates the moments of a random variable that is the number of discrete-steps the
system is in states pertaining to $\Gamma$ during the time interval $t$ \cite{touc09}. Defining the coefficients  
\begin{equation}
\det\left(yI-\mathcal{K}(z)\right)\equiv \sum_{n=0}^{N}c_n(z)y^n,
\label{eqdet2}
\end{equation}
we obtain
\begin{equation}
v_2= -\frac{\sum_{n=0}^{N} c_n'}{\sum_{n=1}^{N}n c_n},
\label{velocity2}
\end{equation}
and
\begin{equation}
D_2= -\frac{1}{2}\lim_{\epsilon\to 0} f(\epsilon),
\label{dispersion2}
\end{equation}
where
\begin{equation}
f(\epsilon)\equiv \epsilon\frac{\sum_{n=0}^{N}c''_n+2v_2\sum_{n=1}^{N}n c'_n+v_2^2\sum_{n=2}^{N}n(n-1)c_n}{\sum_{n=1}^{N}n c_n}.
\label{dispersion3}
\end{equation}
These expressions for velocity and dispersion are derived in appendix \ref{app2}, where we also show that Eqs. (\ref{dispersion2}) and (\ref{dispersion3}) are in perfect agreement with numerical simulations.
The expression for the dispersion of this second class of random variables is the main technical result of this paper. Given a continuous-time Markov process with a small number of states, small in the sense that 
the matrix in Eq. (\ref{modgeneratordisc}) can be handled analytically, one can readily obtain the dispersion
$D_2$ as a function of the transition rates using Eqs. (\ref{dispersion2}) and (\ref{dispersion3}).

In the next section we  calculate the dispersion from different random variables and analyze their relation with energy dissipation. 
Related studies that analyze the relation between the relative uncertainty associated 
with another random variable, the life-time of cluster of states, with energy dissipation are \cite{li02,qian07} for a biological timer, 
and \cite{lang14} for a receptor that dissipates chemical energy. In general, this lifetime of a cluster of states is different from the time an 
stochastic trajectory spends in a cluster of states.

%==========================================================================
\section{Inference of an external ligand concentration by an internal protein}
%==========================================================================
\label{sec4}

%==========================================================================
\subsection{Uncertainty and inference time}
%==========================================================================

\begin{figure}
\includegraphics[width=45mm]{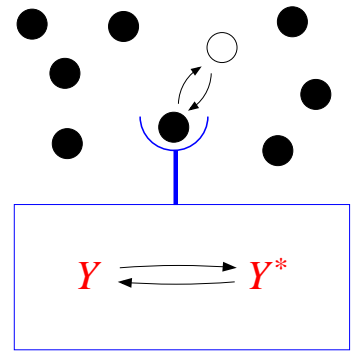}
\vspace{-2mm}
\caption{(Color online) Two-component network. External ligands, represented by the black spheres, can bind to and unbind from the receptor siting on the 
cell membrane. The chemical reaction rates of the protein inside the cell $Y$ depend on the occupancy of the receptor.
}
\label{fig1} 
\end{figure} 

As an application, we consider a two-component network of a cell estimating an external concentration $c$ represented in Fig. \ref{fig1}.
A review on two-component signaling networks is given in \cite{stoc00}. In our model, transition rates can depend on $c$, and, therefore, the moments of a random variable $X$, pertaining 
to any of the two classes, will depend on $c$. The relative uncertainty in estimating $c$ by monitoring $X$ for a time $t$ is written as 
\begin{align}
\left(\frac{(\delta c)^2}{c^2}\right)_X &\equiv \left(c \frac{\partial\langle X\rangle}{\partial c}\right)^{-2}(\langle X^2\rangle-\langle X\rangle^2)\nonumber\\
& = \frac{1}{t} \left(c \frac{\partial v}{\partial c}\right)^{-2}2D\equiv \frac{1}{t}\mathcal{T}.  
\label{uncertaintydef}
\end{align} 
Characteristically, the uncertainty decreases with the integration time $t$. The prefactor $\mathcal{T}$ is the integration time required to get a relative  uncertainty smaller than 100\%. We shall refer to it as the inference time.
Below we will calculate this inference time obtained from different random variables in a two-component network.
The inference time can also be expressed as the inverse of the signal-to-noise ratio normalized by the integration-time time multiplied by the relative change in the
external concentration \cite{skog11,skog13}. 

%==========================================================================
\subsection{Berg and Purcell Limit}
%==========================================================================

\begin{figure}
\includegraphics[width=45mm]{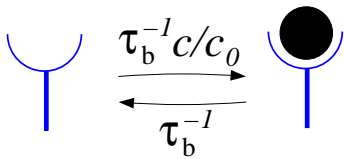}
\vspace{-2mm}
\caption{(Color online) Two-state single receptor model. The bound state corresponds to the receptor with a ligand.
The arrows indicate the transition rates from unbound to bound $\tau_b^{-1}c/c_0$ and from bound to unbound $\tau_b^{-1}$.}
\label{fig2} 
\end{figure}

The receptors sitting on the cell membrane constitute the first layer of the two-component sensory network. In the two-state model for a single receptor shown in Fig. \ref{fig2}, the receptor
can be either bound by a ligand or free. The parameters determining the transition rates in Fig. \ref{fig2} are the external concentration $c$, the concentration for which both states are equally 
probable $c_0$, and the average lifetime of the bound state $\tau_b$. The key feature of these transition rates is that the binding rate is proportional to the external ligand concentration $c$.

If the concentration is estimated by monitoring the time spent in the bound state, the inference time is 
\begin{equation}
\mathcal{T}_{2}=  2\tau_b \frac{c+c_0}{c},  
\label{BPlim}
\end{equation}
which is the result obtained by Berg and Purcell \cite{berg77}. The subscript in $\mathcal{T}_{2}$ indicates that the random variable is from the second class.  

If instead of monitoring the time spent in the bound state the activity, i.e., the total number of transitions is monitored, a
different uncertainty is obtained. In \cite{endr09}, it has been shown that the inference time obtained from the activity is   
\begin{equation}
\mathcal{T}_{1}= \left(1+\frac{c^2}{c_0^2}\right) \tau_b \frac{c+c_0}{c}.  
\label{Actlim}
\end{equation}
For low concentrations $c\ll c_0$, the uncertainty obtained by monitoring the activity is half of the Berg and Purcell uncertainty, i.e., $\mathcal{T}_{1}= \mathcal{T}_{2}/2$. 
This same limit is achieved by estimating the concentration with a maximum likelihood method applied to the probability of a trajectory of 
the two-state model \cite{endr09}. It is straightforward to obtain Eq. (\ref{Actlim}) with the formalism from Sec. \ref{sec2} and Eq. (\ref{BPlim}) with the formalism from Sec. \ref{sec3},
in both cases one has to calculate the characteristic polynomial of a 2$\times$2 matrix.

%==========================================================================
\subsection{Monitoring the internal protein}
%==========================================================================

\begin{figure}
\includegraphics[width=45mm]{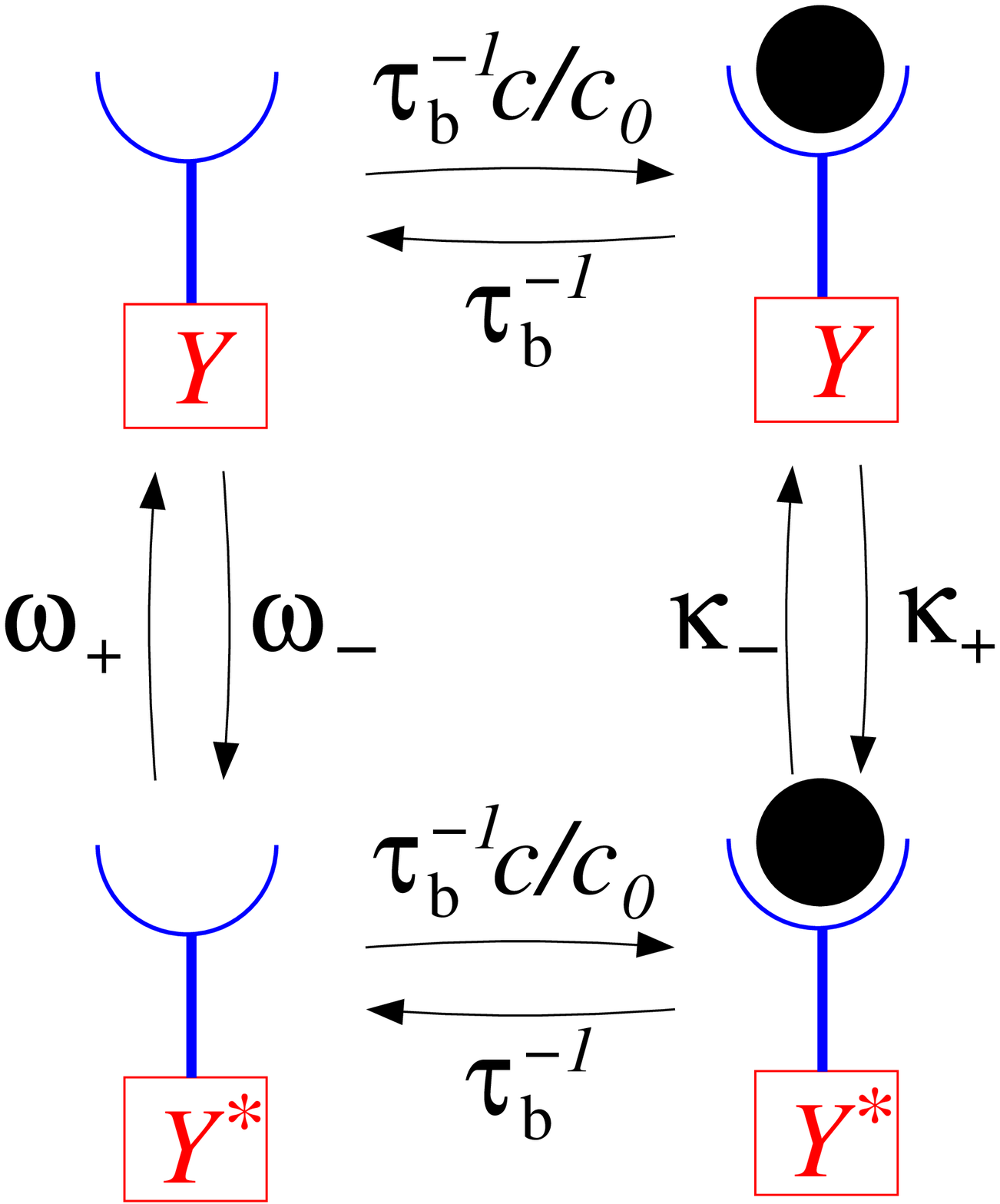}
\vspace{-2mm}
\caption{(Color online) Four-state model accounting for the internal protein $Y$. The transition rates for the internal protein are defined in Eqs. (\ref{eqpho}) and (\ref{eqdepho}). 
}
\label{fig3} 
\end{figure}

The second layer of the two-component network is an internal protein which can be in states $Y$ and $Y^*$, where $Y^*$ is the phosphorylated form of the protein.
The full two-component network then has four states \cite{bara13b}, as shown in Fig. \ref{fig3}. Similar models for two-component networks consider a large 
number of internal proteins \cite{meht12}. If the receptor is bound by a ligand the following phosphorylation reaction 
can happen
\begin{equation}
 Y+ATP\xrightleftharpoons[\kappa_-]{\kappa_+} Y^*+ADP,
\label{eqpho} 
\end{equation}
where $\kappa_{\pm}$ are transition rates. If the receptor is empty, dephosphorylation may take place, i.e.,
\begin{equation}
Y^*\xrightleftharpoons[\omega_-]{\omega_+} Y+P_i.
\label{eqdepho} 
\end{equation}
For thermodynamic consistency these transition rates must fulfill the generalized detailed balance relation \cite{seif12} 
\begin{equation}
\frac{\kappa_+\omega_+}{\kappa_-\omega_-}= \textrm{e}^{\Delta \mu},
\label{eqmu}
\end{equation}
where $\Delta\mu$ is the free energy liberated in one ATP hydrolysis and we are setting $k_BT=1$ throughout. This model is
in equilibrium only if $\Delta \mu=0$, which means that no energy is dissipated. Otherwise, chemical free energy is dissipated,
with the rate of dissipated heat being characterized by the entropy production \cite{seif12}.  

If we are interested in the occupation of the receptor, we obtain the two-state 
model from Fig. \ref{fig2} by integrating $Y$ out since the receptor is independent of the internal protein. 
Hence, the inference time associated with the time the receptor is bound in this four-state model
is as given by Eq. (\ref{BPlim}) and the one associated with the activity of binding/unbinding transitions is as given by Eq. (\ref{Actlim}). 
 
The external concentration can also be estimated by monitoring random variables related to the internal protein. We consider three random variables:
the time the internal protein is in the phosphorylated form, the probability current and the activity of the transitions involving the $Y$ protein.
Note that the present four-state model has only one probability current, which, for $\Delta \mu>0$, goes in the clockwise direction in Fig. \ref{fig3}.
The inference time is denoted by $\mathcal{T}^Y_{2}$ for the time spent in the phosphorylated state. For random variables of the first class we 
denote $\mathcal{T}^Y_{J}$ the inference time related to the current and $\mathcal{T}^Y_{A}$ the one related to the activity.
The calculation of these quantities using the expressions from Sec. \ref{sec2} and Sec. \ref{sec3} is explained in Appendix \ref{app3}. 

%==========================================================================
\subsection{Inference time and dissipation}
%==========================================================================

We first compare  $\mathcal{T}^Y_{J}$ with $\mathcal{T}^Y_{2}$. In equilibrium, the current is zero. Hence, the inference time $\mathcal{T}^Y_{J}$  
diverges for $\Delta \mu=0$. It turns out that also for monitoring the time spent in the phosphorylated state $Y^*$ the inference time $\mathcal{T}^Y_{2}$ diverges in equilibrium.
This divergence comes from the fact that the sensitivity $\partial v^Y_{2}/\partial c$ goes to zero in equilibrium, since the probability of being in the cluster of states 
for which the protein is phosphorylated becomes independent of $c$ for $\Delta\mu=0$.

For the current variable we can obtain the following relation between heat dissipation and inference time. In \cite{bara15} we have obtained a general relation between uncertainty
and dissipated free energy that for the present model reads 
\begin{equation}
2D_J^Y\sigma/[(v_J^Y)^2]\ge 2,
\label{neweq1}
\end{equation}
where $\sigma$ is the thermodynamic entropy production that quantifies the rate of dissipation due to $ATP$ 
consumption. This entropy production is simply given by $\sigma= \Delta \mu v_J^Y$, where $v_J^Y$ is the average current.
From the definition of inference time in Eq. (\ref{uncertaintydef}) and from Eq. (\ref{neweq1}) we obtain
\begin{equation}
\mathcal{T}^Y_{J}\ge   2\left(\frac{\partial \ln v_J^Y}{\partial \ln c}\right)^{-2}\sigma^{-1},
\label{tradeoff}
\end{equation}
where the term $\frac{\partial \ln v_J^Y}{\partial \ln c}$ is the sensitivity of the average current to changes in $c$. Prominently, this lower bound on inference time decreases with
increasing dissipation. This expression provides a tradeoff between inference time, sensitivity and dissipation. Since relation (\ref{neweq1}) is general, the tradeoff relation 
(\ref{tradeoff}) is valid beyond the specific model considered in this section.

For the following calculations we set the transition rates to 
\begin{equation}
\kappa_+=\omega_+= \tau_y^{-1}\textrm{e}^{\Delta \mu/2}\qquad\textrm{and}\qquad\kappa_-=\omega_-= \tau_y^{-1},
\label{choicerates}
\end{equation}
where $\tau_y$ is the time-sale of the reversed transitions in Eqs. (\ref{eqpho}) and (\ref{eqdepho}). We define the dimensionless parameter 
\begin{equation}
\gamma\equiv \tau_y/\tau_b.
\end{equation}

First we take $\gamma\to 0$, which corresponds to the internal protein reactions being fast. The inference time related to the current in this limit is
\begin{equation}
\mathcal{T}^Y_{J}= \frac{(c^2+c_0^2)(\textrm{e}^{\Delta\mu}+1)+4c c_0\textrm{e}^{\Delta\mu/2}}{c_0^2(\textrm{e}^{\Delta\mu/2}-1)^2} \tau_b \frac{c+c_0}{c}.
\end{equation}
For large $\Delta\mu$, this inference time reaches its minimal value given by the uncertainty obtained with the activity of the binding/unbinding transitions $\mathcal{T}_1$ in Eq. (\ref{Actlim}). 
In this case, the phosphorylated form of the protein $Y^*$ is strongly correlated with the 
bound state, while $Y$ is correlated with the unbound state. The binding of ligand is immediately followed by the transition $Y\to Y^*$, whereas the unbinding of a ligand is followed by $Y^*\to Y$. 
Therefore, monitoring the current becomes equivalent to monitoring the activity of the receptor.

\begin{figure}
\includegraphics[width=55mm]{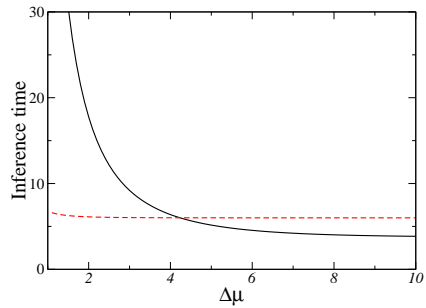}
\vspace{-2mm}
\caption{(Color online) Inference times related to the current $\mathcal{T}^Y_{J}$ (black solid line) and the time spent in the phosphorylated state $\mathcal{T}^Y_{2}$ (red dashed line) 
as a function of $\Delta\mu$ for $\gamma=10^{-2}$, $\tau_b=1$, and $c/c_0=1/2$. 
}
\label{fig4} 
\end{figure}

In the limit $\gamma\to 0$ with any finite $\Delta\mu>0$, the time spent in the phosphorylated state leads to $\mathcal{T}^Y_{2}=\mathcal{T}_2$, 
with $\mathcal{T}_2$ given in Eq. (\ref{BPlim}). While for a large $\Delta\mu>0$ this result can be explained by the fact that $Y^*$ ($Y$) and bound (unbound) receptor become 
strongly correlated it is surprising that for any finite $\Delta\mu>0$ monitoring the internal protein is equivalent to monitoring the receptor.
For both random variables, current and time-spent in the phosphorylated state, the best uncertainty that a variable related 
to the internal protein can reach is bounded by the uncertainty obtained by monitoring a corresponding random variable related to the receptor directly. The main difference between these
two random variables in the present limit is that the precision related to the time spent 
in the phosphorylated state does not improve  by increasing $\Delta \mu$. Hence, there is a region for which increasing energy dissipation leads to an exponential  decrease  
in  $\mathcal{T}^Y_{J}$, while $\mathcal{T}^Y_{2}$ remains approximately constant, as shown in Fig. \ref{fig4}.

\begin{figure}
\includegraphics[width=55mm]{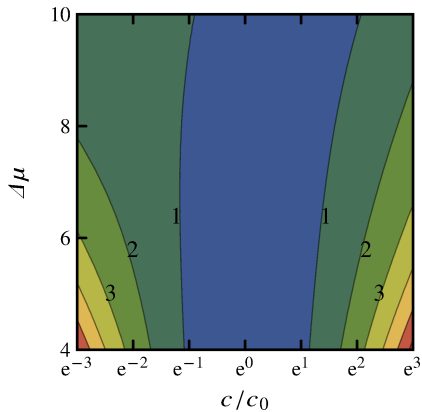}
\vspace{-2mm}
\caption{(Color online) The ratio $\mathcal{T}^Y_{2}/\mathcal{T}^Y_{J}$ in the $\ln(c/c_0)\times \Delta\mu$ plane for $\gamma=100$ and $\tau_b=1$. 
The region in blue corresponds to $\mathcal{T}^Y_{2}/\mathcal{T}^Y_{J}<1$, where monitoring the current leads to a smaller inference time compared to monitoring the time
spent in the phosphorylated state.
}
\label{fig5} 
\end{figure}

Another important limit corresponds to the case where the reversed transitions of the internal protein are much slower than unbinding of the ligand, i.e.,  $\gamma\to \infty$.
In this limit, where the relevant time-scale for the averaging time required for a small uncertainty is $\tau_y\gg \tau_b$, 
the inference times related to current and time spent in the phosphorylated state are
\begin{equation}
\mathcal{T}^Y_{J}= \frac{(c + c_0)^2 [4 c c_0 \textrm{e}^{\Delta\mu/2} + (c^2+ c_0^2)(\textrm{e}^{\Delta\mu}+1)]}{c (c - c_0)^2 c_0 (\textrm{e}^{\Delta\mu/2}-1)^2 (\textrm{e}^{\Delta\mu/2}+1)}\tau_y+\textrm{O}(\tau_b)
\label{sy1jinf}
\end{equation}
and
\begin{equation}
\mathcal{T}^Y_{2}= \frac{2 (c + c_0)^2 [(c^2+c_0^2)\textrm{e}^{\Delta\mu/2} + c c_0 (\textrm{e}^{\Delta\mu}+1)] }{c^2 c_0^2 (\textrm{e}^{\Delta\mu/2}-1)^2 (\textrm{e}^{\Delta\mu/2}+1)}\tau_y +\textrm{O}(\tau_b),
\label{sy2inf}
\end{equation}
respectively. In these expressions we are assuming that the prefactors are large compared to $\tau_b/\tau_y$. Particularly, for large $\Delta\mu$ both the prefactor in Eq. (\ref{sy1jinf}) and 
the prefactor in Eq. (\ref{sy2inf}) decay with $\textrm{e}^{-\Delta\mu/2}$. In this case, the expressions remain valid for $\textrm{e}^{-\Delta\mu/2}\gg \tau_b/\tau_y$.  
Comparing Eq. (\ref{sy1jinf}) with Eq. (\ref{sy2inf}) we see that for $c$ close to $c_0$  $\mathcal{T}^Y_{J}\gg \mathcal{T}^Y_{2}$, while  for $c\gg c_0$ or $c\ll c_0$, $\mathcal{T}^Y_{J}\ll \mathcal{T}^Y_{2}$. 
Hence for intermediate values of the concentration close to $c_0$ the inference
time related to the current is larger. This feature is quantified in Fig. \ref{fig5}, showing that the ratio $\mathcal{T}^Y_{2}/\mathcal{T}^Y_{J}$ 
is smaller than $1$ in an intermediate concentration range. The precise range for which monitoring the time spent in the phosphorylated state leads to a 
smaller uncertainty than monitoring the current depends on $\Delta \mu$.
     
\subsection{Dissipationless inference}

Finally, we analyze the inference time arising from monitoring the activity of the transitions involving changes in the internal protein $Y$. 
For the transition rates set in Eq. (\ref{choicerates}) we observe that $\mathcal{T}^Y_{A}\ge\mathcal{T}^Y_{J}$.
Furthermore, in the limit of $\gamma\to 0$ and large $\Delta \mu$, for which $Y^*$ and bound receptor are strongly correlated,
the inference time  $\mathcal{T}^Y_{A}=\mathcal{T}^Y_{J}=\mathcal{T}_{1}$, with $\mathcal{T}_{1}$ given by Eq. (\ref{Actlim}).

The main feature of the inference time $\mathcal{T}^Y_{A}$ is that it can be finite even in equilibrium. Therefore, while the other
two variables related to $Y$, namely, the current and the time spent in the phosphorylated state, require dissipation to infer an external
concentration, by monitoring the activity it is possible to infer the concentration without any dissipation. In order to make this feature 
transparent we set the transition rates as      
\begin{equation}
\kappa_+=\kappa_-= a\tau_y^{-1}\qquad\textrm{and}\qquad\omega_+=\omega_-= \tau_y^{-1},
\label{choicerates2}
\end{equation}
which from Eq. (\ref{eqmu}) imply $\Delta \mu=0$. The factor $a>0$ quantifies how much faster the transitions
that change $Y$ are when the receptor is bound. If $a>1$ internal transitions are faster with the receptor bound and if
$a<1$ they are faster for a free receptor. The inference time with the rates in Eq. (\ref{choicerates2}) becomes
\begin{equation}
\mathcal{T}^Y_{A}= \frac{(c + c_0)[(a-1)^22cc_0^2+\gamma B] )}{c^2 c_0^2 (a-1)^2}\tau_b,
\end{equation}  
where $B\equiv  a(c^3+2c^2 c_0+cc_0^2)+(c_0^3+c^2 c_0+2cc_0^2)$. For large $a$, $\mathcal{T}^Y_{A}\to \mathcal{T}_{2}$, with $\mathcal{T}_{2}$ given by Eq. (\ref{BPlim}). 
This result can be explained as follows. For $a\gg 1$, whenever the receptor is bound a large number of internal transitions take place while for a unbound receptor this
number is much smaller. Hence, counting the number of transitions of the internal protein, i.e., the activity of the internal protein, becomes equivalent to monitoring the time spent in the bound state.
The surprising feature here is that depending on the random variable, it is possible to reach the Berg and Purcell limit by monitoring the internal protein with no energy dissipation.

%==========================================================================
\subsection{Comparison to related work}
%==========================================================================

Let us compare our results with recent work on two-component signaling networks that estimate an external concentration \cite{meht12,gove14,gove14a}. In these works the random variable is 
the fluctuating number of internal proteins that are in the phosphorylated (or active) state, which can be viewed as memory of the receptor occupancy. As demonstrated in \cite{meht12}, 
the uncertainty related to this random variable diverges in equilibrium, similar to current and time spent in the phosphorylated state. A main point of the present paper is that the relation between dissipation 
and uncertainty depends on which random variable we choose to estimate the external concentration, as manifested by the fact that activity allows for dissipationless inference. We note that 
for a different model with internal proteins directly binding to free receptors the uncertainty estimated from the number of internal proteins can also be finite in equilibrium \cite{gove14a}.

The rate of mutual information in bipartite systems, for which the two-component network is an example, has been studied in \cite{bara13b,bara13a}. In these 
references it has been shown that the rate of mutual information between an external process, which corresponds to the receptor in our model, and an internal process, corresponding to the protein $Y$, 
can be non-zero even in equilibrium. A posteriori, this result indicates that even in equilibrium the inference of an external concentration with finite uncertainty should be possible if an appropriate random variable is monitored. 
We have shown that the activity is a random variable that achieves this dissipationless inference.

%==========================================================================
\section{Conclusion}
%==========================================================================
\label{sec5}

We have developed a general formalism that allows for the calculation of the dispersion of the time a stochastic trajectory spends in a cluster of 
states that can be applied to arbitrary Markov processes with a finite number of states. This method is potentially useful for any application 
where the dispersion of this random variable is of interest. Furthermore, our formalism is similar to the formalism to calculate the dispersion
for another class of random variables comprising current and activity, which has been pioneered by Koza \cite{koza99}. The expressions in Sec. \ref{sec2} and Sec. \ref{sec3}
thus provide a unified recipe to calculate the dispersion of these two classes of random variables.

These expressions have been used in the four-state model for a two-component network estimating an external ligand concentration in Sec. \ref{sec4}. 
With this application, we have compared the uncertainties of the external concentration that are obtained by monitoring three different random variables: current, activity of
the internal protein, and the time the internal protein is in the phosphorylated state. The main results obtained with our case study are as follows.
The best performance (smallest uncertainty) that a random variable related to the internal protein can achieve is limited by random variables associated 
with the receptor, which is the first layer of the network. This best performance is achieved when the internal protein is much faster than the receptor.
In general, determining which random variable leads to the smallest uncertainty depends on parameters like the external concentration and the chemical potential driving 
the phosphorylation cycle. If the internal protein is much slower than the receptors, the integration time to obtain a reliable estimate from the internal 
protein must be much larger than the typical time of a transition of the protein, which is much larger than the lifetime of the bound state of the receptor.
Finally, while current and time spent in the phosphorylated state require energy dissipation for a finite uncertainty, it is possible to estimate the
external concentration from the activity of the internal protein even in equilibrium, which we call dissipationless inference. 
%This result is reminiscent of the fact that the rate of mutual information between the receptor and the internal protein can be non-zero in equilibrium \cite{bara13b,bara13a}.

Recently, we have obtained a universal bound on the energetic cost of the relative uncertainty of a generic probability current \cite{bara15}. As shown here,
this bound implies a general bound on the inference time based on a current variable. An intriguing question is whether it is
possible to find such universal relations between energy dissipation and uncertainty for other random variables like the activity or the time a trajectory spends in a cluster of states.
The formalism developed in this paper provides a tool to investigate the existence of such thermodynamic uncertainty relations that connect energy dissipation with uncertainty.

\appendix

\section{Derivation of expressions for the first class}
\label{app1}

The maximum eigenvalue of the modified generator in Eq. (\ref{modgenerator}) is denoted $\lambda(z)$. It can be shown that this maximum eigenvalue
is the scaled cumulant generating function associated with the random variable $X_1$ \cite{lebo99,touc09}. The velocity 
and dispersion are related to $\lambda(z)$  as
\begin{equation}
v_1= \lambda'
\label{Jcum}
\end{equation}
and
\begin{equation}
D_1= \lambda''/2,
\label{dcum}
\end{equation}
respectively. The lack of explicit $z$ dependence denotes evaluation at $z=0$.  
Since $\lambda(z)$ is a root of the polynomial in Eq. (\ref{eqdet}), it follows that
\begin{equation}
\sum_{n=0}^{N}C_n(z)[\lambda(z)]^n=0.
\label{poly2}
\end{equation}
Using the fact that $\lambda(0)=0$, we obtain Eqs. (\ref{eqJC}) and (\ref{eqDC}) from Eqs. (\ref{Jcum}) and (\ref{dcum}) by taking
the first and second derivatives with respect to $z$ in Eq. (\ref{poly2}), respectively. These calculations can be found in more detail
in \cite{bara15,bara15a}, for example.

\section{Derivation of expressions for the second class}
\label{app2}

The time interval $t$ is discretized in $L$ intervals of size $\epsilon$, with $t= L\epsilon$. The time evolution of the corresponding Markov chain is determined by
the discrete-time stochastic matrix in Eq. (\ref{discsto}). We consider a random variable $x_\epsilon$ which counts in how many
of the $L$ segments the system was in the cluster of states $\Gamma$. We denote $\psi(z)$ the maximum eigenvalue of the matrix in 
Eq. (\ref{modgeneratordisc}). Hence, $\psi(z)$ is a root of the polynomial in Eq. (\ref{eqdet2}), i.e.,
\begin{equation}
\sum_{n=0}^{N}c_n(z)[\psi(z)]^n=0.
\label{poly3}
\end{equation}
For $z=0$, the matrix in Eq. (\ref{modgeneratordisc}) becomes the discrete-time stochastic matrix in Eq. (\ref{discsto}), which implies
$\psi(0)=1$. By taking first and second derivatives in Eq. (\ref{poly3}) and setting $z=0$ we obtain
\begin{equation}
\psi'=-\frac{\sum_{n=0}^{N}c_n'}{\sum_{n=1}^{N}nc_n}
\label{psip1}
\end{equation}
and
\begin{equation}
\psi''=-\frac{\sum_{n=0}^{N}c_n''+2\psi'\sum_{n=1}^{N}nc_n'+(\psi')^2\sum_{n=2}^{N}n(n-1)c_n}{\sum_{n=1}^{N}nc_n}.
\label{psipp1}
\end{equation}
Furthermore this maximum eigenvalue is the generating function associated with $x_\epsilon$ \cite{touc09}, i.e., 
\begin{equation}
\ln\psi(z)=\frac{1}{L}\ln\langle \textrm{e}^{z x_\epsilon} \rangle.
\end{equation}
Hence, in the large $L$ limit,
\begin{equation}
\psi'= \frac{\langle x_\epsilon\rangle}{L} 
\label{psip2}
\end{equation}
and
\begin{equation}
\psi''= \frac{\langle x_\epsilon^2\rangle-\langle x_\epsilon\rangle^2}{L}. 
\label{psipp2}
\end{equation}

\begin{figure}
\subfigure[]{\includegraphics[width=30mm]{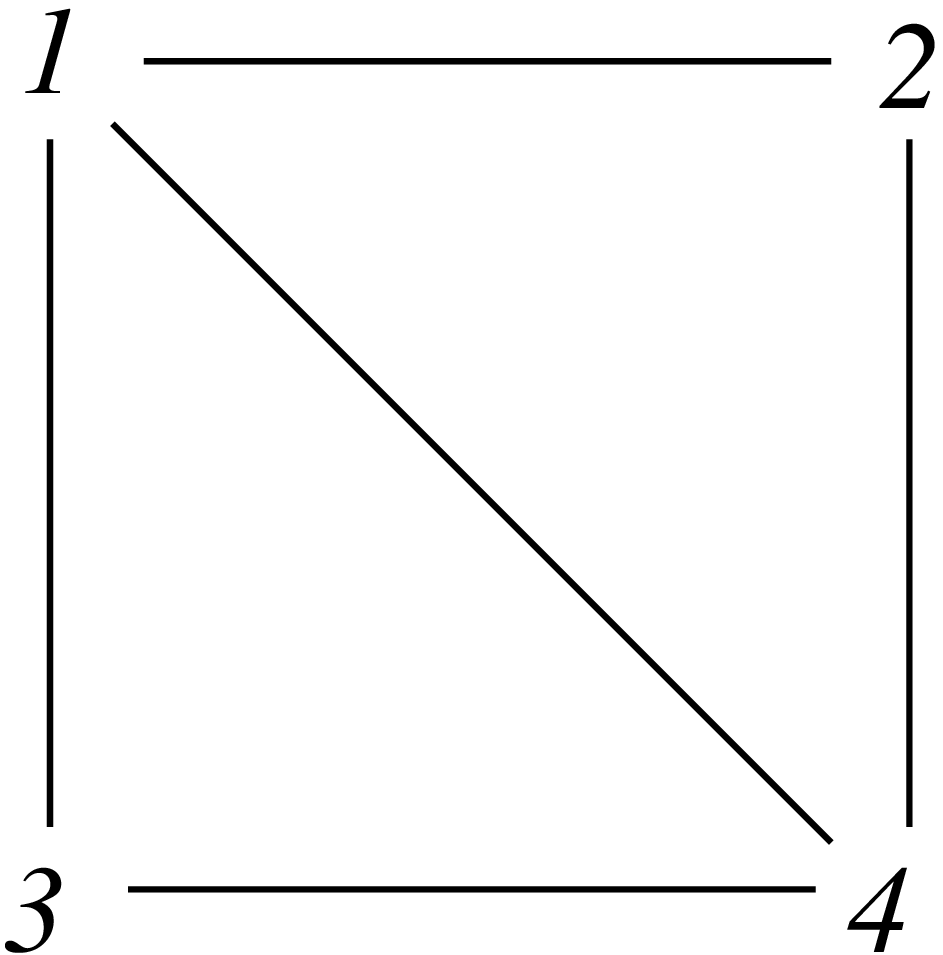}\label{fig6a}}\hfill
\subfigure[]{\includegraphics[width=30mm]{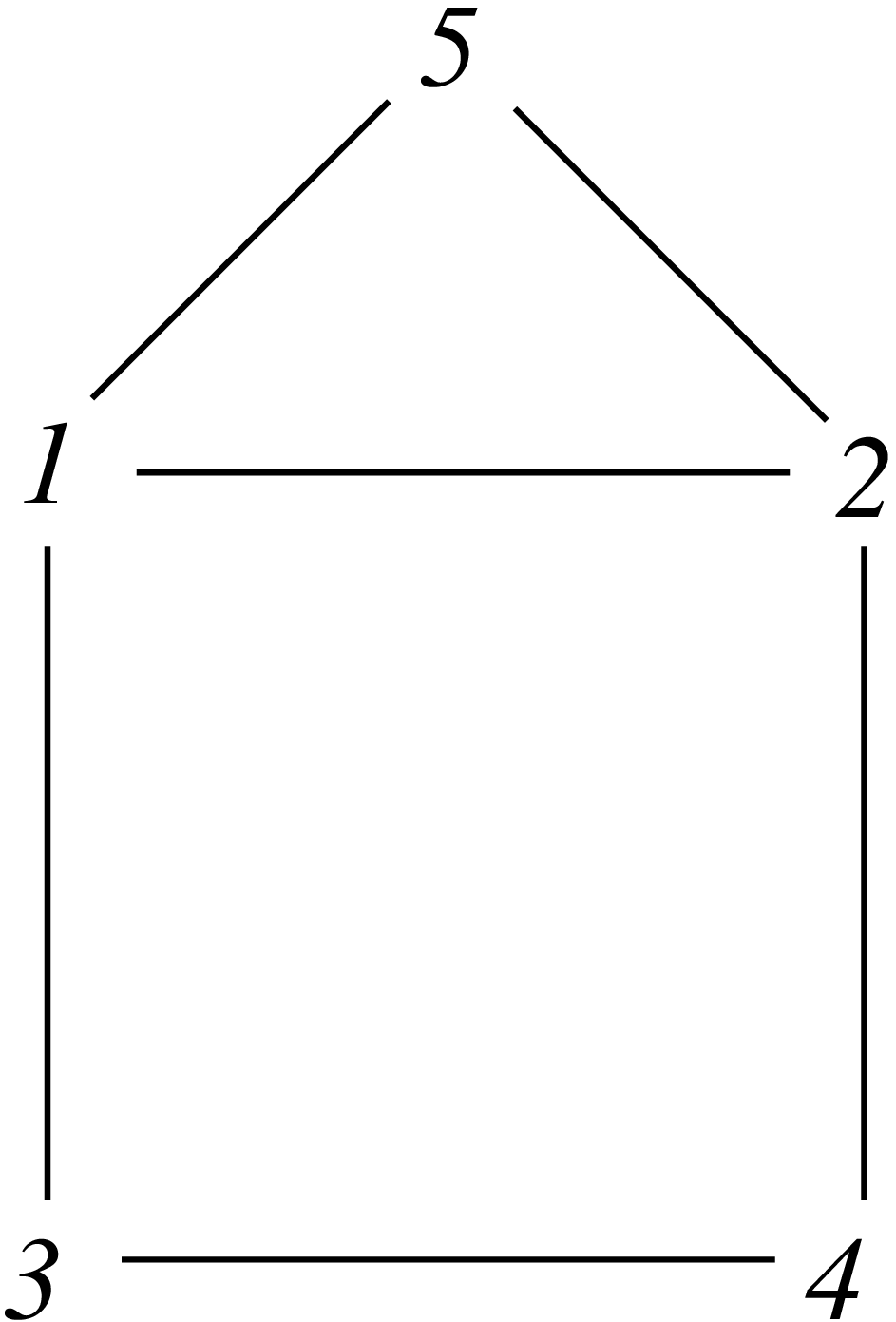}\label{fig6b}}
\vspace{-2mm}
\caption{Networks for which numerical tests have been performed. The numbers represent states and the links
represent transition rates that are not zero. (a) Network with four states. Note that for $w_{14}=w_{41}=0$ this network of
states becomes the same as the network from Fig. \ref{fig3}. (b) Network with five states.
}
\label{fig6} 
\end{figure}

\begin{figure}
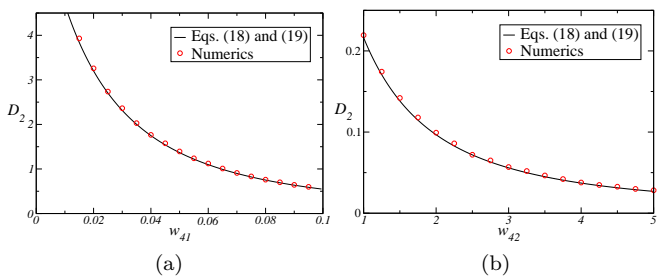

\subfigure[]{\includegraphics[width=43mm]{./0num4state2cyc.eps}}\hfill
\subfigure[]{\includegraphics[width=43mm]{./05states.eps}}
\vspace{-2mm}
\caption{Dispersion $D_2$ obtained from Eqs. (\ref{dispersion2}) and (\ref{dispersion3}) compared to numerical simulations for:
(a) Network from Fig. \ref{fig6a} where the time spent in state $4$ is the random variable and the rates are $w_{12}=10$, $w_{13}=2$, $w_{31}=8$, and $w_{14}=w_{21}=w_{24}=w_{42}=w_{43}=w_{34}=0.01$.
(b) Five states network from Fig. \ref{fig6b} where the time spent in the cluster $\Gamma=(1,4)$ is the random variable and the rates are $w_{43}=20$, $w_{34}=8.001$, $w_{31}=0.1$, $w_{13}=0.005$, $w_{12}=2$,
$w_{15}=7$, $w_{24}=9.9$, $w_{21}=4$, $w_{25}=57$, $w_{51}=0.02$, and $w_{52}=0.6$.
}
\label{fig7} 
\end{figure}

The first cumulant of the discrete and continuous variables fulfill the relation 
\begin{equation}
\frac{\langle X_2 \rangle}{t}=\frac{\langle x_\epsilon \rangle}{L}.
\label{eq1app3} 
\end{equation}
This relation comes from the fact that the stationary distribution of the continuous-time Markov process is the the same as the discrete-time
Markov chain. The second cumulant of the continuous variable is related to the second cumulant of $x_\epsilon$ by 
\begin{equation}
\frac{\langle X_2^2 \rangle-\langle X_2^2 \rangle}{t}=\lim_{\epsilon\to 0}\epsilon\frac{\langle x_\epsilon^2 \rangle-\langle x_\epsilon \rangle^2}{L}. 
\label{eq2app3}
\end{equation}  
The factor $\epsilon$ on the right hand side accounts for the correct dimension. This relation is equivalent to $\langle X_2^2 \rangle/t^2= \lim_{\epsilon\to 0}\langle x_\epsilon^2 \rangle/L^2$.
Note that in Eqs. (\ref{eq1app3}) and (\ref{eq2app3}) the average in the left hand side is 
over continuous-time trajectories, whereas the average in the right hand
side is over discrete-time trajectories. From Eqs. (\ref{psip1}), (\ref{psip2}) and (\ref{eq1app3}) we obtain Eq. (\ref{velocity2}). 
From Eqs. (\ref{psipp1}), (\ref{psipp2}) and (\ref{eq2app3}) we obtain Eq. (\ref{dispersion2}), with $f(\epsilon)$ given by Eq. (\ref{dispersion3}).

Even though we do not have a rigorous proof for Eq. (\ref{eq2app3}) we have verified numerically its validity for several different networks of states by comparing the dispersion obtained 
from continuous-time numerical simulations with the dispersion obtained from our analytical expressions in Eqs. (\ref{dispersion2}) and (\ref{dispersion3}). For example, we have performed 
this test for the networks of states in Fig. \ref{fig6} with the results shown in Fig. \ref{fig7}.

\section{Calculations for the four-state system}
\label{app3}
The four states of the model in Fig. \ref{fig3} are identified as $(Y,u)\mathrel{\hat=} 1$, $(Y,b)\mathrel{\hat=} 2$, $(Y^*,u)\mathrel{\hat=} 3$, and $(Y^*,b)\mathrel{\hat=} 4$, where $b$ ($u$) denotes bound (unbound) receptor.
The transition rates are given by $w_{12}=w_{34}= \tau_b^{-1}c/c_0$, $w_{21}=w_{43}= \tau_b^{-1}$, $w_{13}=\omega_-$, $w_{31}=\omega_+$, $w_{24}=\kappa_+$, and $w_{42}=\kappa_-$. 
This four-state model has only one probability current, which is the same in any of the four links in Fig. \ref{fig3}. If we take the current from $1$ to $2$, the respective modified generator in Eq. \ref{modgenerator}
is 
\begin{equation}
\mathcal{L}_J(z)\equiv\left(
\begin{array}{cccc}
-r_1 & w_{21} \textrm{e}^{-z} & w_{31} & 0 \\
 w_{12}\textrm{e}^z          & -r_2 & 0 & w_{42} \\ 
 w_{13}          & 0 & -r_3 & w_{43} \\
0          & w_{24} & w_{34}& -r_4
\end{array}
\right),  
\label{matrix1}
\end{equation} 
where $r_1= w_{12}+w_{13}$, $r_2= w_{21}+w_{24}$, $r_3= w_{31}+w_{34}$, and $r_4= w_{42}+w_{43}$. The subscript $J$ indicates the modified generator for the current. In the case of the activity of the transitions changing the internal
portein state, which are transitions involving the pairs $13$ and $24$, the modified generator reads
\begin{equation}
\mathcal{L}_A(z)\equiv\left(
\begin{array}{cccc}
-r_1 & w_{21}  & w_{31}\textrm{e}^{z} & 0 \\
 w_{12}          & -r_2 & 0 & w_{42}\textrm{e}^{z} \\ 
 w_{13}\textrm{e}^{z}          & 0 & -r_3 & w_{43} \\
0          & w_{24}\textrm{e}^{z} & w_{34}& -r_4
\end{array}
\right). 
\label{matrix2}
\end{equation} 
The inference times $\mathcal{T}^Y_{J}$ and $\mathcal{T}^Y_{A}$ are calculated from the coefficients of the characteristic polynomial in Eq. (\ref{eqdet}) with expressions (\ref{eqJC}), (\ref{eqDC}), and (\ref{uncertaintydef}), 
where for $\mathcal{T}^Y_{J}$ the modified generator is in Eq. (\ref{matrix1}) and for $\mathcal{T}^Y_{A}$ the modified generator is in Eq. (\ref{matrix2}). In both cases, we can get full analytical expressions for arbitrary 
transition rates, however they are too long to be displayed here. 

For the time spent in the phosphorylated state of the protein, the corresponding cluster is $\Gamma=(3,4)$. The matrix in Eq. (\ref{modgeneratordisc}) then becomes
\begin{equation}
\mathcal{K}(z)\equiv\left(
\begin{array}{cccc}
(1-r_1\epsilon) & w_{21}\epsilon  & w_{31}\epsilon & 0 \\
 w_{12}\epsilon          & (1-r_2\epsilon) & 0 & w_{42}\epsilon \\ 
 w_{13}\epsilon\textrm{e}^{z}        & 0 & (1-r_3\epsilon)\textrm{e}^{z} &  w_{43}\epsilon\textrm{e}^{z} \\
0          & w_{24}\epsilon\textrm{e}^{z} & w_{34}\epsilon\textrm{e}^{z} & (1-r_4\epsilon)\textrm{e}^{z}
\end{array}
\right).  
\label{matrix3}
\end{equation} 
With the coefficients of the characteristic polynomial associated with this matrix, as given by Eq. (\ref{eqdet2}), the inference time  $\mathcal{T}^Y_{2}$ is obtained from Eqs. (\ref{velocity2}), (\ref{dispersion2}), (\ref{dispersion3}) 
and (\ref{uncertaintydef}). We also get a long analytical expression for $\mathcal{T}^Y_{2}$ in terms of the transition rates.

%\begin{acknowledgments}
%Support by the ESF though the network EPSD  is gratefully acknowledged. 
%\end{acknowledgments}
%==========================================================================
% References
%==========================================================================

\end{document}